\documentclass[runningheads]{llncs}
\usepackage{svg}
\usepackage{caption}
\usepackage{subcaption}
\usepackage{amsmath, amssymb}
\usepackage[skins,breakable]{tcolorbox}
\usepackage{hyperref}
\usepackage{siunitx}
\usepackage{multicol}
\usepackage{multirow}
\usepackage{xcolor}
\usepackage{booktabs}
\definecolor{real}{HTML}{4363d8}
\definecolor{simulated}{HTML}{911eb4}
\definecolor{UNIT}{HTML}{e6194B}
\definecolor{cINN}{HTML}{00CC96}

\usepackage{graphicx}

\begin{document}
\title{Unsupervised Domain Transfer with Conditional Invertible Neural Networks}
\author{Kris K. Dreher\inst{1, 2}\thanks{Send correspondence to K.K.D. k.dreher@dkfz-heidelberg.de or L.M.H. l.maier-hein@dkfz-heidelberg.de} \and
Leonardo Ayala\inst{1, 3} \and
Melanie Schellenberg\inst{1, 4} \and
Marco Hübner\inst{1, 4} \and
Jan-Hinrich Nölke\inst{1, 4} \and
Tim J. Adler\inst{1} \and
Silvia Seidlitz\inst{1, 4, 5, 6} \and
Jan Sellner\inst{1, 4, 5} \and
Alexander Studier-Fischer\inst{7} \and
Janek Gröhl\inst{8, 9} \and
Felix Nickel\inst{7} \and
Ullrich Köthe\inst{3} \and
Alexander Seitel\inst{1} \and
Lena Maier-Hein\inst{1, 3, 4, 5, 6, \star}
}
\authorrunning{K.K. Dreher et al.}

\institute{
Intelligent Medical Systems, German Cancer Research Center (DKFZ), Heidelberg, Germany \and
Heidelberg University, Faculty of Physics and Astronomy, Heidelberg, Germany \and
Medical Faculty, Heidelberg University, Heidelberg, Germany \and
Faculty of Mathematics and Computer Science, Heidelberg University, Heidelberg, Germany \and
Helmholtz Information and Data Science School for Health, Karlsruhe/Heidelberg, Germany \and
National Center for Tumor Diseases (NCT) Heidelberg, a partnership between DKFZ and Heidelberg University Hospital, Heidelberg, Germany \and
Department of General, Visceral, and Transplantation Surgery, Heidelberg University Hospital, Heidelberg, Germany\and
Cancer Research UK Cambridge Institute, University of Cambridge, Cambridge, United Kingdom \and
Department of Physics, University of Cambridge, Cambridge, United Kingdom
}

\maketitle              %
\begin{abstract}
Synthetic medical image generation has evolved as a key technique for neural network training and validation. A core challenge, however, remains in the domain gap between simulations and real data. While deep learning-based domain transfer using Cycle Generative Adversarial Networks and similar architectures has led to substantial progress in the field, there are use cases in which state-of-the-art approaches still fail to generate training images that produce convincing results on relevant downstream tasks. Here, we address this issue with a domain transfer approach based on conditional invertible neural networks (cINNs). As a particular advantage, our method inherently guarantees cycle consistency through its invertible architecture, and network training can efficiently be conducted with maximum likelihood training. To showcase our method’s generic applicability, we apply it to two spectral imaging modalities at different scales, namely hyperspectral imaging (pixel-level) and photoacoustic tomography (image-level). According to comprehensive experiments, our method enables the generation of realistic spectral data and outperforms the state of the art on two downstream classification tasks (binary and multi-class). cINN-based domain transfer could thus evolve as an important method for realistic synthetic data generation in the field of spectral imaging and beyond.

\keywords{Domain transfer  \and invertible neural networks \and medical imaging \and photoacoustic tomography \and hyperspectral imaging \and deep learning.}
\end{abstract}

\section{Introduction}
\label{sec:introduction}

\begin{figure}[h]
    \centering
    \includegraphics{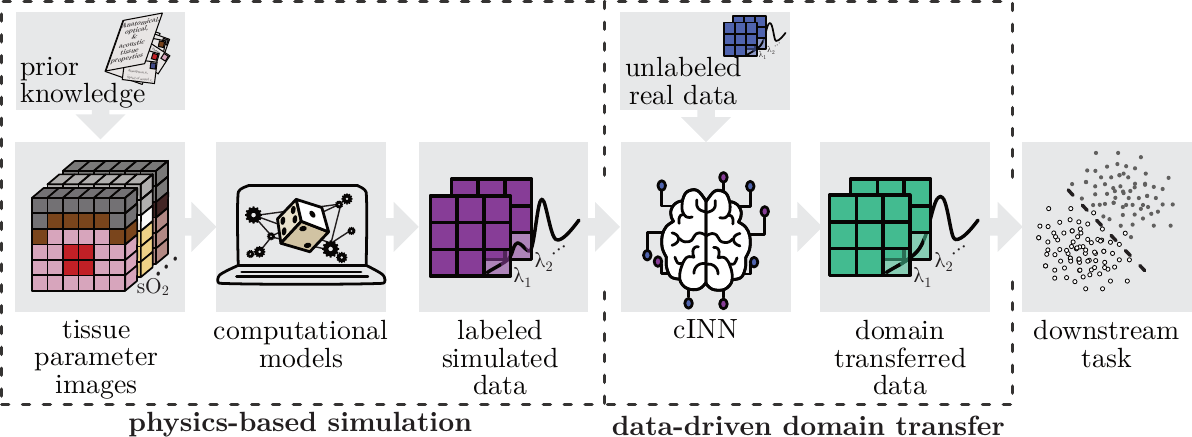}
    \caption{\textbf{Pipeline for data-driven spectral image analysis in the absence of labeled reference data.} A physics-based simulation framework generates simulated spectral images with corresponding reference labels (e.g., tissue type or oxygenation (sO$_2$)). Our domain transfer method based on cINNs leverages unlabeled real data to increase their realism. The domain-transferred data can then be used for supervised training of a downstream task (e.g. classification). 
 }
    \label{fig:concept}
\end{figure}

The success of supervised learning methods in the medical domain led to countless breakthroughs that might be translated into clinical routine and have the potential to revolutionize healthcare \cite{de2018clinically,isensee2021nnu}. For many applications, however, labeled reference data (ground truth) may not be available for training and validating a neural network in a supervised manner. One such application is spectral imaging which comprises various non-interventional, non-ionizing imaging techniques that can resolve functional tissue properties such as blood oxygenation in real time \cite{adler2019uncertainty,Ayala2022,Ayala_2019,wirkert_physiological_2017}. 
While simulations have the potential to overcome the lack of ground truth, synthetic data is not yet sufficiently realistic~\cite{grohl2021deep}. 
Cycle Generative Adversarial Networks (GAN)-based architectures are widely used for domain transfer~\cite{barth2020optimising,hoffman2018cycada} but may suffer from issues such as unstable training, hallucinations, or mode collapse \cite{li2020adversarial}. Furthermore, they have predominantly been used for conventional RGB imaging and one-channel cross-modality domain adaptation, and may not be suitable for other imaging modalities with more channels.
We address these challenges with the following contributions:\\ 
\textbf{Domain transfer method}: We present an entirely new sim-to-real transfer approach based on conditional invertible neural networks (cINNs) (cf. Fig. \ref{fig:concept}). Our architecture features inherent cycle consistency and the possibility of conducting maximum likelihood learning while still maintaining the high visual quality of adversarial networks, without the possibility of mode collapse. \\
\textbf{Instantiation to spectral imaging}: We show that our method can generically be applied to two complementary modalities: photoacoustic tomography (PAT; image-level) and hyperspectral imaging (HSI; pixel-level).\\
\textbf{Comprehensive validation}: In comprehensive validation studies based on more than 2,000 PAT images (real: $\sim$ 1,000) and more than 6 million spectra for HSI (real: $\sim$ 6 million) we investigate and subsequently confirm our two main hypotheses: (H1) Our cINN-based models can close the domain gap between simulated and real spectral data better than current state-of-the-art methods regarding spectral plausibility. (H2) Training models on data transferred by our cINN-based approach can improve their performance on the corresponding (clinical) downstream task without them having seen labeled real data.

\section{Materials and Methods}
\label{sec:Materials_and_Methods}

\subsection{Domain Transfer with Conditional Invertible Neural Networks}
\label{sec:DomainTransfer_with_cINN}

\begin{figure}[h]
    \centering
    \includegraphics{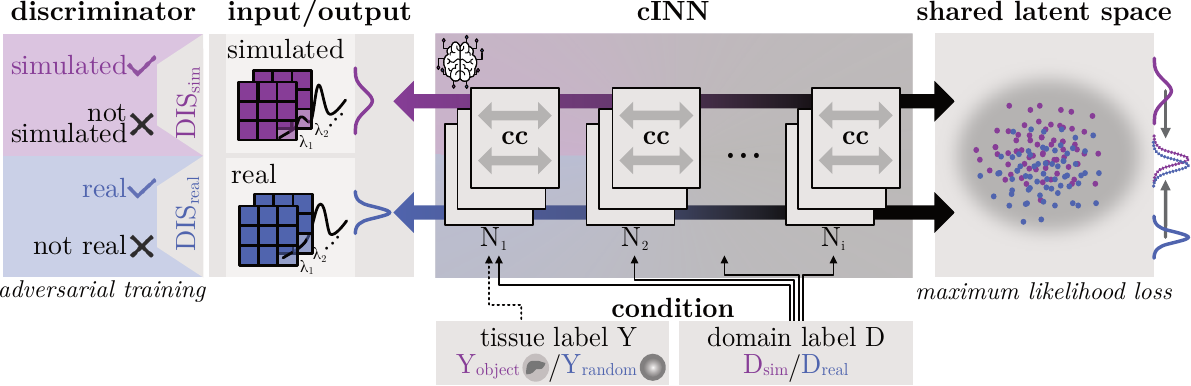}
    \caption{\textbf{Proposed architecture based on cINNs.} The invertible architecture transfers both simulated and real data into a shared latent space (right). By conditioning on the domain \textit{D} (bottom), a latent vector can be transferred to either the simulated or the real domain (left) for which the discriminator $\text{Dis}_\text{sim}$ and $\text{Dis}_\text{real}$ calculate the losses for adversarial training.
 }
    \label{fig:INN}
\end{figure}
\noindent
\textbf{Concept overview. }
Our domain transfer approach (cf. Fig. \ref{fig:INN}). It is based on the assumption that data samples from both domains carry domain-invariant information (e.g., on optical tissue properties) and domain-variant information (e.g., modality-specific artifacts). The invertible architecture, which inherently guarantees cycle consistency, transfers both simulated and real data into a shared latent space. While the domain-invariant features are captured in the latent space, the domain-variant features can either be filtered (during encoding) or added (during decoding) by utilizing a domain label \textit{D}. The additional tissue label \textit{Y} for simulated data implicitly carries information to aid the spectral consistency, whereas the randomly generated proxy label for the unlabeled real data does not. The joint distribution is learned using the maximum likelihood loss and by adding two multiscale discriminators $Dis_{sim}$ and $Dis_{real}$, adversarial training ensures high visual quality of the generated data.

\textbf{Model design. }
The proposed cINN (cf. Fig. \ref{fig:INN}) is roughly based on the work of Ardizzone et. al.~\cite{ardizzone2020conditional} and  consists of multiple ($i$) scales of $N_i$-chained affine conditional coupling (CC) blocks \cite{dinh2016density}. These scales are necessary in order to increase the receptive field of the network and are achieved by Haar wavelet downsampling \cite{haar1911theorie}. 
A CC block consists of subnetworks that can be freely chosen depending on the data dimensionality (e.g., fully connected or convolutional networks) as they are only evaluated in the forward direction. 
The CC blocks receive a condition consisting of two parts: domain label and tissue label, which are then concatenated to the input along the channel dimension. In the case of PAT, the tissue label is a full semantic and random segmentation map for the simulated and real data, respectively. In the case of HSI, the tissue label is a one-hot encoded vector for organ labels.

\textbf{Model training. }
In the following, the proposed cINN with its parameters $\theta$ will be referred to as $f(x, DY,\theta)$ and its inverse as $f^{-1}$ for any input $x\sim p_D$ from domain $D\in\{D_{sim}, D_{real}\}$ with prior density $p_D$ and its corresponding latent space variable $z$. The condition $DY$ is the combination of domain label $D$ as well as the tissue label $Y\in\{Y_{sim}, Y_{real}\}$. Then the maximum likelihood loss $\mathcal{ML}$ for a training sample $x_i$ is described by 
\begin{equation}
    \underset{D}{\mathcal{ML}}=\mathbb{E}_{i}\left[ \frac{||f(x_i,D Y,\theta)|| _2^2}{2} - log |J_i| \right] \text{ with } J_i=det\left(\left.\frac{ \partial f }{\partial x  } \right|_{x_i}\right).
\end{equation}
For the adversarial training, we employ the least squares training scheme \cite{mao2017least} for generator $Gen_D=f^{-1}_D\circ f_{D'}$ and discriminator $Dis_D$ for each domain with $x_{D’}$ as input from the source domain and $x_D$ as input from the target domain:
\begin{equation}
    \underset{{Gen}_D}{\mathcal{L}}=\underset{x_D'\sim p_{D'}}{\mathbb{E}}\left[ (Dis_D(Gen_D(x_{D'})-1))^2\right]
\end{equation}
\begin{equation}
    \underset{{Dis}_D}{\mathcal{L}}=\underset{x_D\sim p_D}{\mathbb{E}}\left[ (Dis_D(x_D)-1)^2\right] + \underset{x_D'\sim p_D'}{\mathbb{E}}\left[ (Dis_D(Gen_D(x_{D'})))^2\right].
\end{equation}
Finally, the full loss for the proposed model comprises the following:
\begin{equation}
    \underset{Total_{Gen}}{\mathcal{L}}=\underset{real}{\mathcal{ML}} + \underset{sim}{\mathcal{ML}} + \underset{{Gen}_{real}}{\mathcal{L}} + \underset{{Gen}_{sim}}{\mathcal{L}} \text{ and }
        \underset{Total_{Dis}}{\mathcal{L}}=\underset{{Dis}_{real}}{\mathcal{L}} + \underset{{Dis}_{sim}}{\mathcal{L}}.
\end{equation}

\textbf{Model inference.}
\label{sec:ModelInference}
The domain transfer is done in two steps: 1) A simulated image is encoded in the latent space with conditions $D_{sim}$ and $Y_{sim}$ to its latent representation $z$, 2) $z$ is decoded to the real domain via $D_{real}$ with the simulated tissue label $Y_{sim}$: $x_{sim \rightarrow real}= f^{-1}(\cdot, D_{real}Y_{sim},\theta) \circ f(\cdot, D_{sim}Y_{sim},\theta)(x_{sim}).$

\begin{figure}[h]
    \centering
    \includegraphics{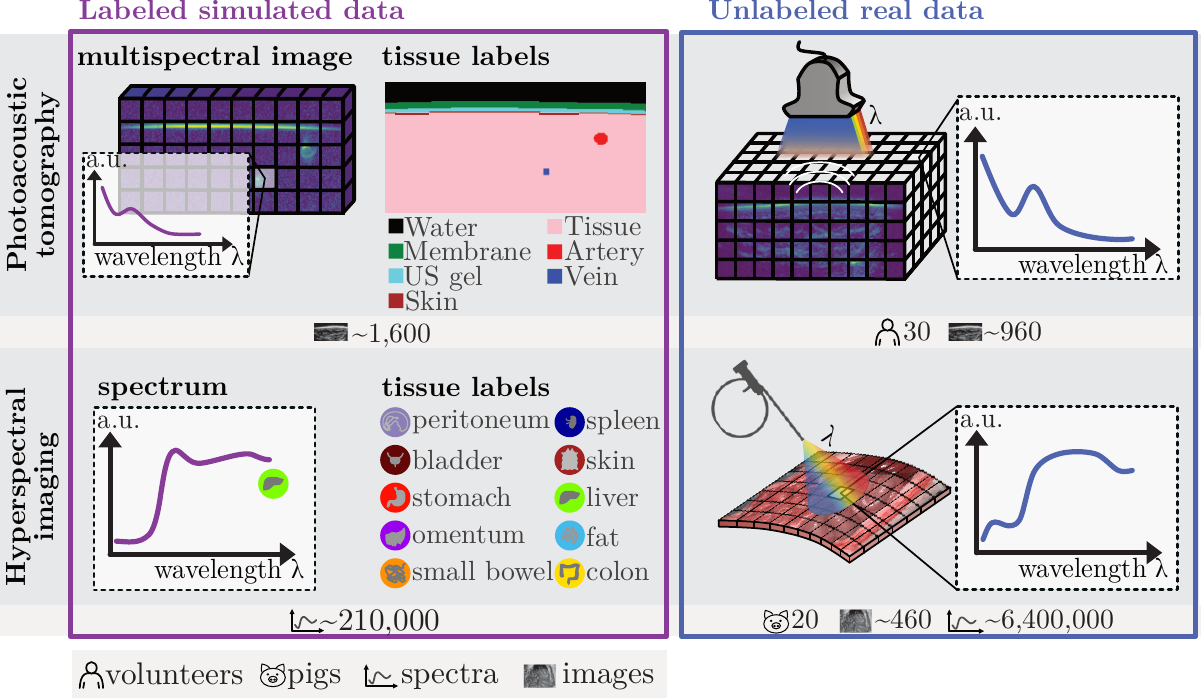}
    \caption{\textbf{Training data used for the validation experiments.} For PAT, ~960 real images from 30 volunteers were acquired. For HSI, more than six million spectra corresponding to 460 images and 20 individuals were used. The tissue labels PAT correspond to 2D semantic segmentations, whereas the tissue labels for HSI represent 10 different organs. For PAT,  $\sim$~1600 images were simulated, whereas around 210,000 spectra were simulated for HSI.
 }
    \label{fig:Data}
\end{figure}

\subsection{Spectral Imaging Data}
\label{sec:SpectralImagingData}

\textbf{Photoacoustic tomography data. }
PAT is a non-ionizing imaging modality that enables the imaging of functional tissue properties such as tissue oxygenation~\cite{wang2006noninvasive}. The \textbf{real PAT data} (cf. Fig. \ref{fig:Data}) used in this work are images of human forearms that were recorded from 30 healthy volunteers using the MSOT Acuity Echo (iThera Medical GmbH, Munich, Germany) (all regulation followed under study ID: S-451/2020, and the study is registered with the German Clinical Trials Register under reference number DRKS00023205). In this study, 16 wavelengths from 700 nm to 850 nm in steps of 10 nm were recorded for each image. 
The resulting 180 images were semantically segmented into the structures shown in Fig. \ref{fig:Data} according to the annotation protocol provided in \cite{schellenberg2021}. Additionally, a full sweep of each forearm was performed to generate more unlabeled images, thus amounting to a total of 955 real images. 
The \textbf{simulated PAT data} (cf. Fig. \ref{fig:Data}) used in this work comprises 1,572 simulated images of human forearms. They were generated with the toolkit for Simulation and Image Processing for Photonics and Acoustics (SIMPA) \cite{grohl2022simpa} based on a forearm literature model \cite{schellenberg2022photoacoustic} and with a digital device twin of the MSOT Acuity Echo. \\
\textbf{Hyperspectral imaging data. }
HSI is an emerging modality with high potential for surgery~\cite{clancy2020surgical}. In this work, we performed pixel-wise analysis of HSI images.
The \textbf{real HSI data} was acquired with the Tivita\textsuperscript{\textregistered} Tissue (Diaspective Vision GmbH, Am Salzhaff, Germany) camera, featuring a spectral resolution of approximately 5 nm in the spectral range between 500 nm and 1000 nm. In total, 458 images, corresponding to 20 different pigs, were acquired (all regulations followed under study IDs: 35-9185.81/G-161/18 and 35-9185.81/G-262/19) and annotated with ten structures: bladder, colon, fat, liver, omentum, peritoneum, skin, small bowel, spleen, and stomach (cf. Fig. \ref{fig:Data}). This amounts to 6,410,983 real spectra in total. 
The \textbf{simulated HSI data} was generated with a Monte Carlo method (cf. algorithm provided in the supplementary material).
This procedure resulted in 213,541 simulated spectra with annotated organ labels.

\section{Experiments and Results}
The purpose of the experiments was to investigate hypotheses H1 and H2 (cf. Sec.~\ref{sec:introduction}). As state-of-the-art method for the experiments, an unsupervised image-to-image translation (UNIT) network~\cite{liu2017unsupervised} in its original version (fully convolutional) and an adapted version for the one-dimensional HSI data was implemented. To make the comparison fair, the tissue label conditions were concatenated with the input. 
\begin{figure}[ht!]
    \centering
    \includegraphics{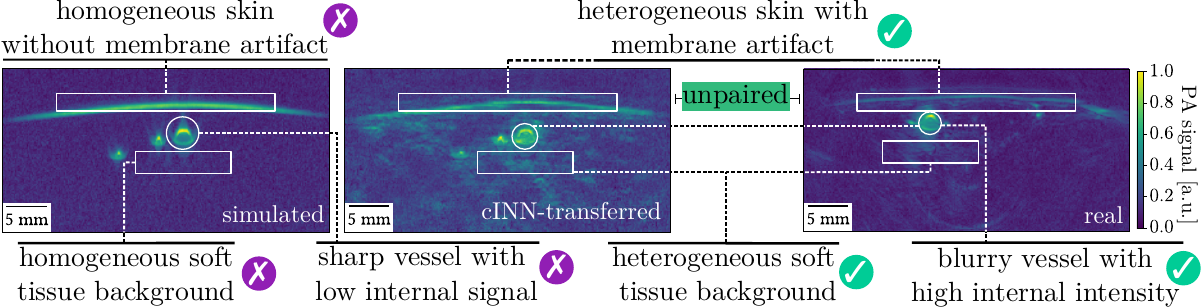}
    \caption{\textbf{Qualitative results.} In comparison to simulated PAT images (left), images generated by the cINN (middle) resemble real PAT images (right) more closely. All images show a human forearm at 800 nm.}
    \label{fig:pai_qualitative_comparison}
\end{figure}

\begin{figure}[h]
    \centering
    \includegraphics[width=\textwidth]{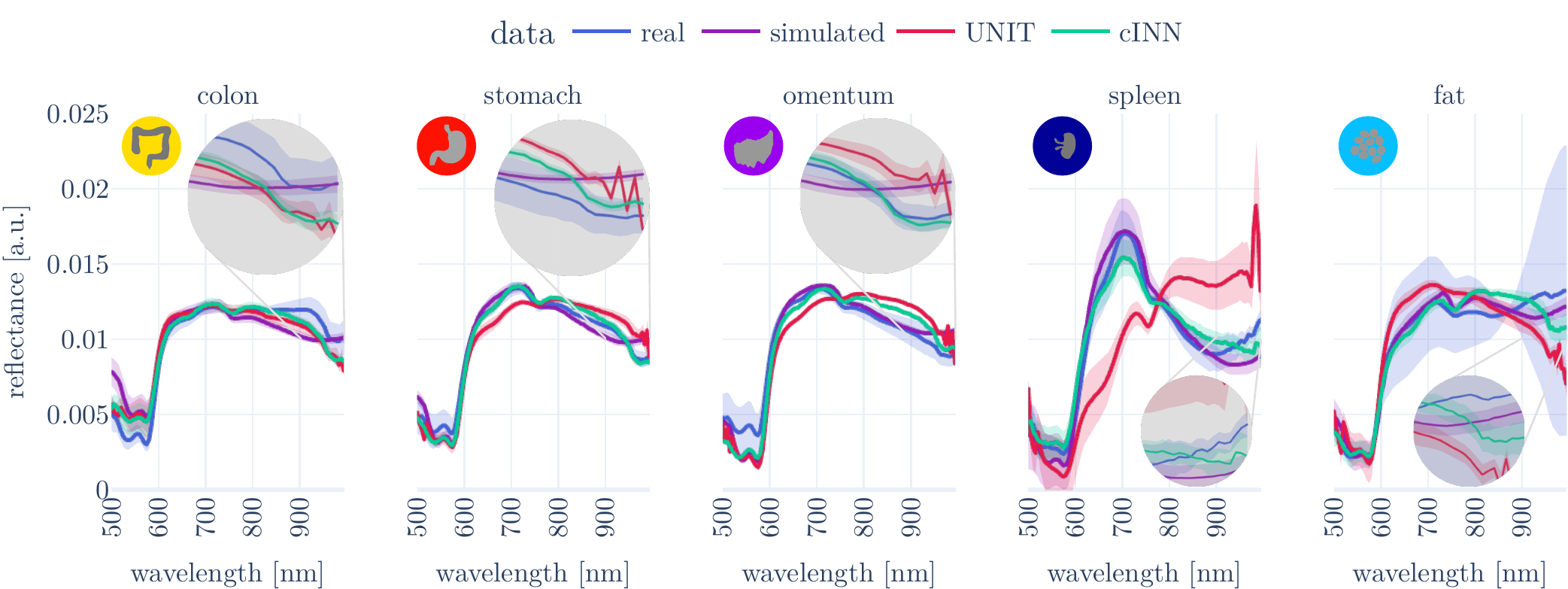}
    \caption{\textbf{The cINN-transferred spectra are in closer agreement with the real spectra than the simulations and the UNIT-transferred spectra}. Spectra for five exemplary organs are shown from 500 nm to 1000 nm. For each subplot, a zoom-in for the near-infrared region ($>$ 900 nm) is shown. The translucent bands represent the standard deviation across spectra for each organ.
}
    \label{fig:semantic_reflectance}
\end{figure}

\textit{Realism of synthetic data (H1)}: According to qualitative analyses (Fig. \ref{fig:pai_qualitative_comparison}) our domain transfer approach improves simulated PAT images with respect to key properties, including the realism of skin, background, and sharpness of vessels. 
\begin{figure}[hb!] 
    \centering
    \includegraphics[width=0.9\textwidth]{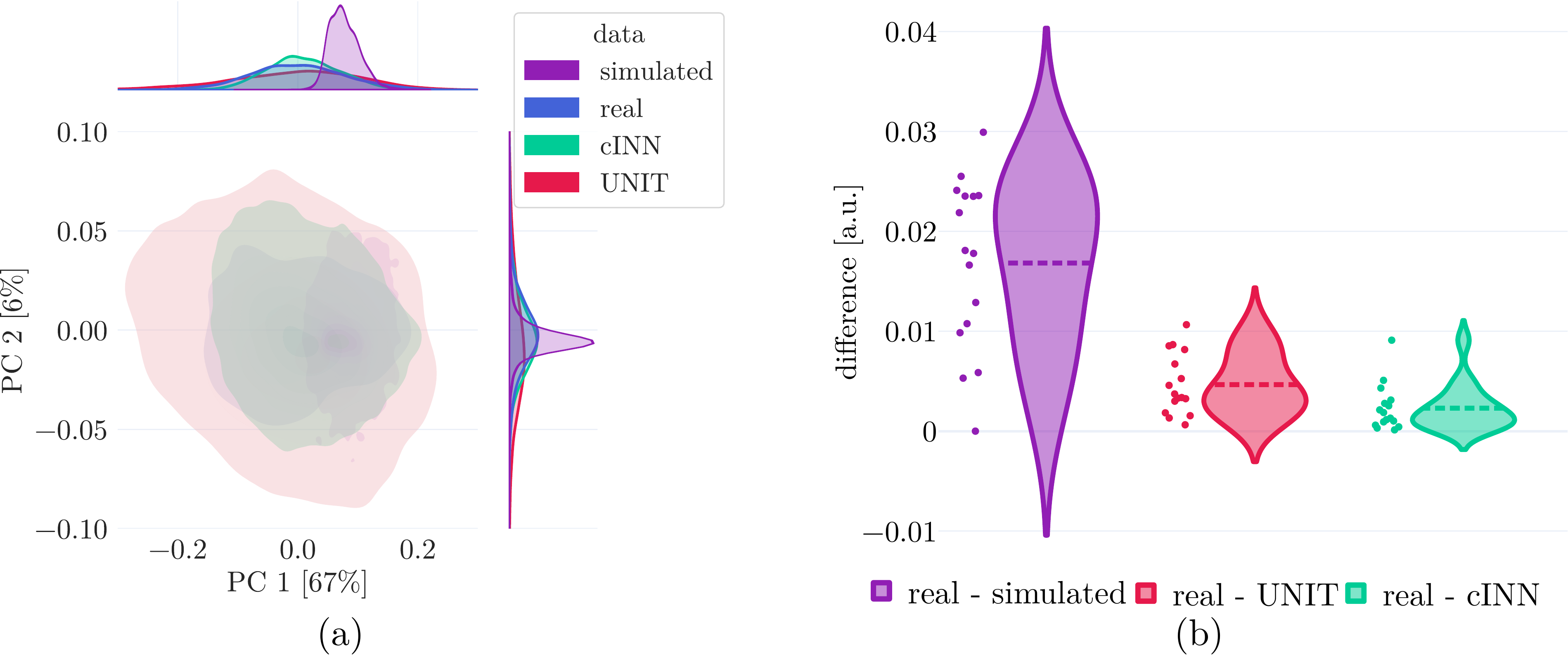}
    \caption{\textbf{Our domain transfer approach yields realistic spectra (here: of veins)}. The PCA plots in a) represent a kernel density estimation of the first and second components of a PCA embedding of the real data, which represent about 67\% and 6\% of the variance in the real data, respectively. The distributions on top and on the right of the PCA plot correspond to the marginal distributions of each dataset’s first two components. b) Violin plots show that the cINN yields spectra that feature a smaller difference to the real data compared to the simulations and the UNIT-generated data. The dashed lines represent the mean difference value, and each dot represents the difference for one wavelength.}
    \label{fig:pai_pca_diff}
\end{figure}\\
A principal component analysis (PCA) performed on all artery and vein spectra of the real and synthetic datasets demonstrates that the distribution of the synthetic data is much closer to the real data after applying our domain transfer approach (cf. Fig. \ref{fig:pai_pca_diff} a)). The same holds for the absolute difference, as shown in Fig. \ref{fig:pai_pca_diff} b). Slightly better performance was achieved with the cINN compared to the UNIT. Similarly, our approach improves the realism of HSI spectra, as illustrated in Fig. \ref{fig:semantic_reflectance}, for spectra of five exemplary organs (colon, stomach, omentum, spleen, and fat). The cINN-transferred spectra generally match the real data very closely. Failure cases where the real data has a high variance (translucent band) are also shown.\\

\textit{Benefit of domain-transferred data for downstream tasks (H2)}:
We examined two classification tasks for which reference data generation was feasible: classification of veins/arteries in PAT and organ classification in HSI. For both modalities, we used the completely untouched real test sets, comprising 162 images in the case of PAT and $\sim$~920,000 spectra in the case of HSI. For both tasks, a calibrated random forest classifier (sklearn \cite{scikit-learn} with default parameters) was trained on the simulated, the domain-transferred (by UNIT and cINN), and real spectra. As metrics, the balanced accuracy (BA), area under receiver operating characteristic (AUROC) curve, and F1-score were selected based on \cite{metrics_reloaded}. 
\begin{figure}[h] 
    \centering
    \includegraphics[width=\textwidth]{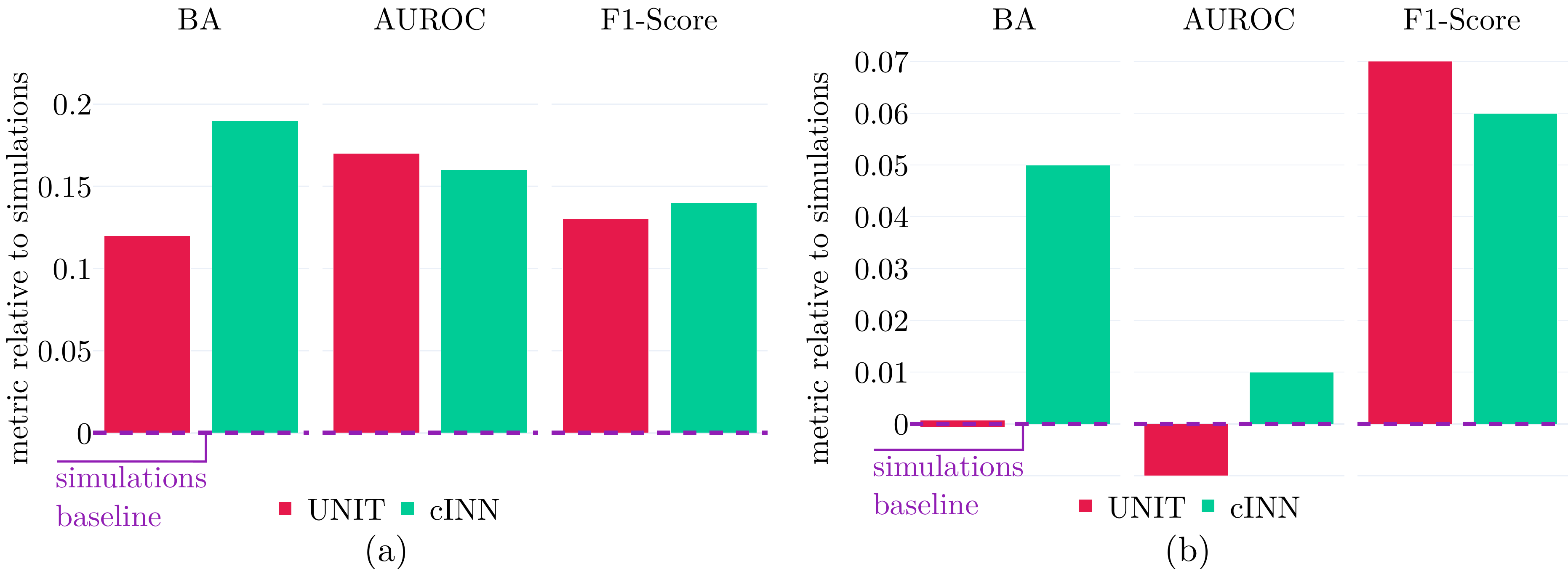}
    \caption{\textbf{Both domain transfer methods increase the classification performance compared to simulated data, but the cINN outperforms the UNIT}. BA, AUROC, and F1-score values were weighted and aggregated to account for class imbalance for both a) artery-vein classification in the case of PAT and b) organ classification in the case of HSI. The zero-baseline corresponds to the reference values of the simulated data.}
    \label{fig:metrics}
\end{figure}

As shown in Fig. \ref{fig:metrics}, our domain transfer approach dramatically increases the classification performance for both downstream tasks. Compared to physics-based simulation, the cINN obtained a relative improvement of 37\% (BA), 25\% (AUROC), and 22\% (F1 Score) for PAT whereas the UNIT only achieved a relative improvement in the range of 20\%-27\% (depending on the metric). For HSI, the cINN achieved a relative improvement of 21\% (BA), 1\% (AUROC), and 33\% (F1 Score) and it scored better in all metrics except for the F1 Score than the UNIT. For all metrics, training on real data still yields better results (see supplementary material). 

\section{Discussion}

With this paper, we presented the first domain transfer approach that combines the benefits of cINNs (exact maximum likelihood estimation) with those of GANs (high image quality). A comprehensive validation involving qualitative and quantitative measures for the remaining domain gap and downstream tasks suggests that the approach is well-suited for sim-to-real transfer in spectral imaging. For both PAT and HSI, the domain gap between simulations and real data could be substantially reduced, and a dramatic increase in downstream task performance was obtained - also when compared to the popular UNIT approach.

The only similar work on domain transfer in PAT has used a cycle GAN-based architecture on a single wavelength with only photon propagation as PAT image simulator instead of full acoustic wave simulation and image reconstruction \cite{li2022deep}. This potentially leads to spectral inconsistency in the sense that the spectral information either is lost during translation or remains unchanged from the source domain instead of adapting to the target domain.
Outside the spectral/medical imaging community, Liu et al. \cite{liu2017unsupervised} and Grover et al. \cite{grover2020alignflow} tasked variational autoencoders and invertible neural networks for each domain, respectively, to create the shared encoding. They both combined this approach with adversarial training to achieve high-quality image generation. Das et al. \cite{das2021cdcgen} built upon this approach by using labels from the source domain to condition the domain transfer task. In contrast to previous work, which used en-/decoders for each domain, we train a single network as shown in Fig. \ref{fig:INN}. with a two-fold condition consisting of a domain label ($D$) and a tissue label ($Y$) from the source domain, which has the advantage of explicitly aiding the spectral domain transfer.

The main limitation of our approach is the high dimensionality of the parameter space of the cINN as dimensionality reduction of data is not possible due to the information and volume-preserving property of INNs. This implies that the method is not suitable for arbitrarily high dimensions. Future work will comprise the rigorous validation of our method with tissue-mimicking phantoms for which reference data are available. 

In conclusion, our proposed approach of cINN-based domain transfer is a novel method enabling the generation of realistic spectral data. As it is not limited to spectral data, it could develop into a powerful method for domain transfer in the absence of labeled real data for a wide range of image modalities in the medical domain and beyond.

\setcounter{page}{1}
\setcounter{figure}{0}
\renewcommand{\thefigure}{S\arabic{figure}}
\renewcommand{\thetable}{S\arabic{table}}

\section*{Supplementary Material for: Unsupervised Domain Transfer with Conditional Invertible Neural Networks}

\begin{table}[h!]
\centering
\caption{\textbf{Simulated ranges of physiological parameters of a three-layer  tissue model for HSI}. $v_{\textrm{Hb}} \lbrack \% \rbrack$ represents the blood volume fraction, $sO_2$ the blood oxygenation, $a_{mie}$ the reduced scattering coefficient at $500$\,nm, $b_{mie}$ the scattering power, $g$ the scattering anisotropy, $n$ the refractive index, and $d$ the layer thickness. Parameters were uniformly sampled within the specified range.}
\label{tab:generic}
\begin{tabular}{p{1.8cm} c c c c c c c}
\hline
      & $v_{\textrm{Hb}} \lbrack \% \rbrack$           & $sO_2 \lbrack \% \rbrack$  & $a_{\text{mie}} \lbrack \si{cm^{-1}} \rbrack$ & $b_\text{mie}$[a.u.]         & $g$[a.u.]      & $n$[a.u.]         & $d \lbrack\si{cm}\rbrack$      \\ 
\hline
layers 1 to 3: & $0-30$ & $0-100$ & $5-50$ & $0.3-3$ & $0.80-0.95$ & $1.33-1.54$ & $0.002-0.2$  \\ 
\hline \hline
\multicolumn{8}{l}{simulation framework: GPU-MCML, $10^6$ photons per simulation}                                                \\                  
\multicolumn{8}{l}{simulated samples: \num{5.5e5} in  wavelength range: 500\,\si{nm}-1000\,\si{nm}, step size 2\,\si{nm}}\\
\hline
\end{tabular}
\end{table}

\begin{figure}[h]
    \centering
    \includegraphics[width=0.9\textwidth]{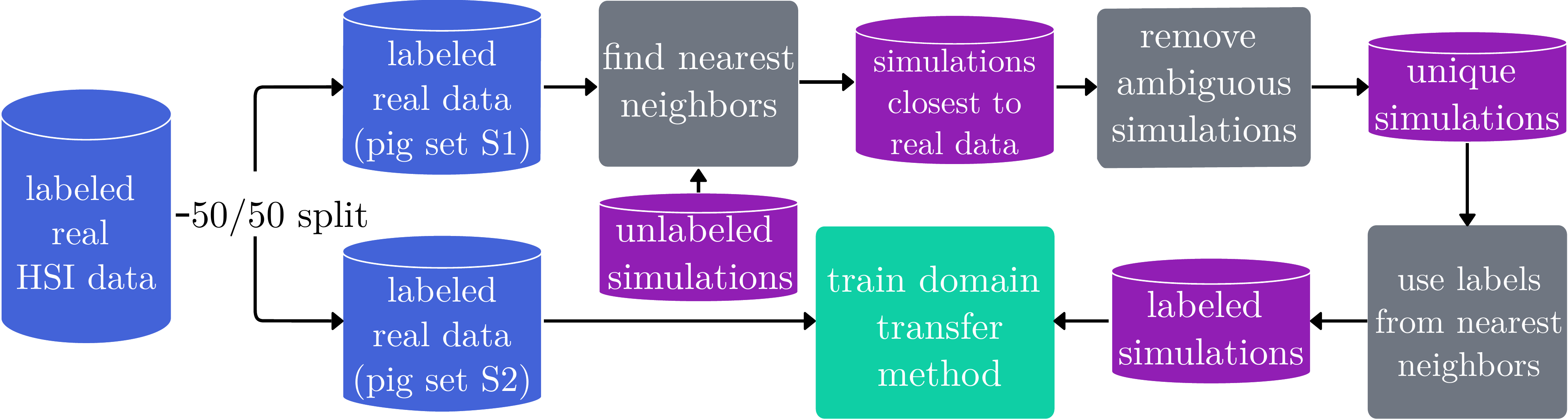}
    \caption{\textbf{HSI data generation}.   Real data is leveraged to reduce a large set of simulations to a set of plausible spectra with unambiguous labels.}
    \label{fig:suppl_knn_data_generation}
\end{figure}

\begin{figure}[hb!]
    \centering
    \includegraphics[width=\textwidth]{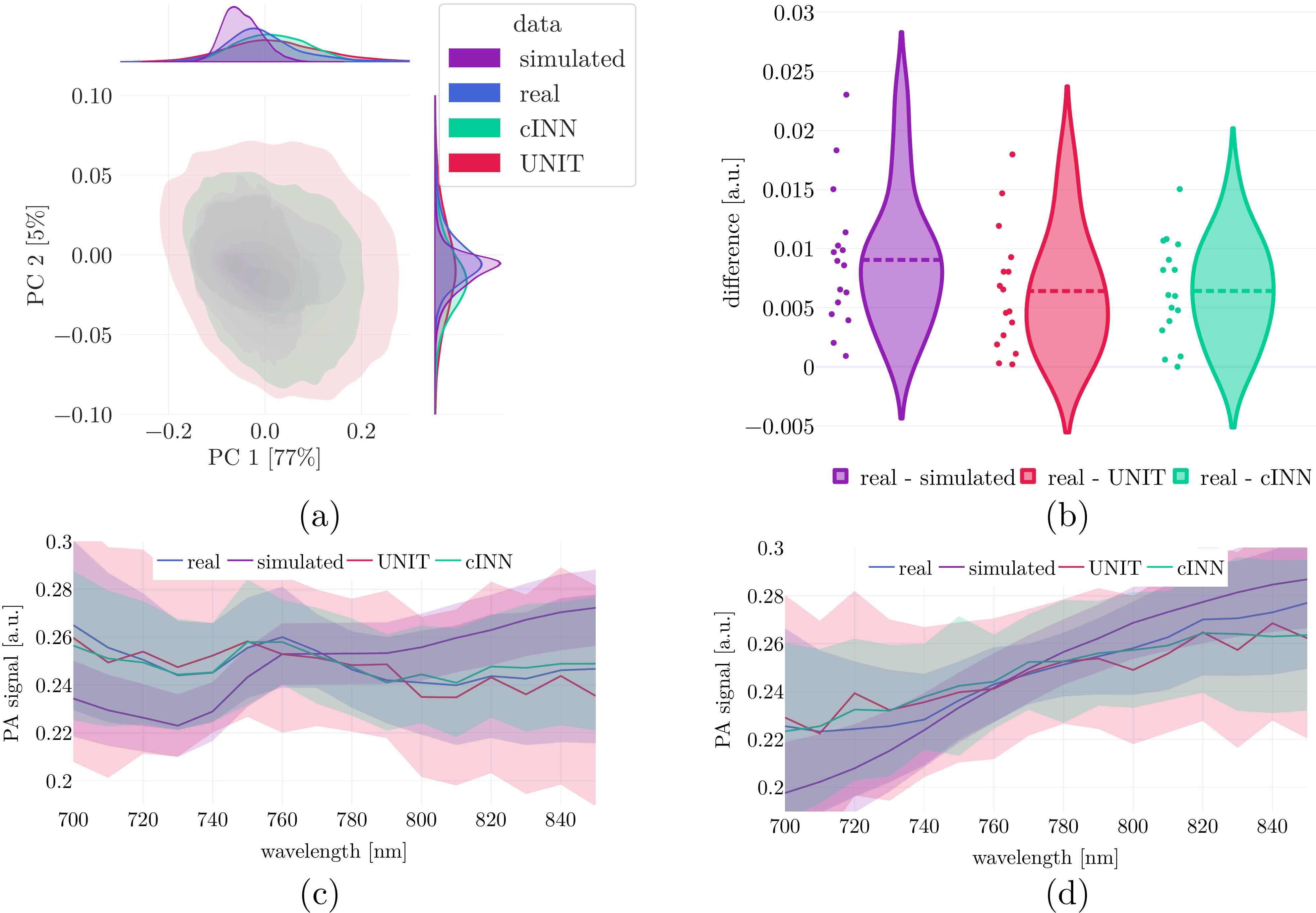}
    \caption{\textbf{Our domain transfer approach yields realistic spectra}. Figures a) and b) are the artery-equivalent of veins in Fig. 5 of the main paper. c) and d) represent the real, simulated, and domain-transferred vein and artery spectra.}
    \label{fig:suppl_artery}
\end{figure}

\begin{table}[hb!]
\centering
\caption{\textbf{Classification scores for different training data.} The training data refers to real data, simulated data without domain transfer, data generated by a UNIT with ($\text{UNIT}_\text{Y}$) and without ($\text{UNIT}$) tissue labels, and by a cINN with (proposed $\text{cINN}_\text{DY}$) and without ($\text{cINN}_\text{D}$) tissue labels as condition, respectively. The best performing methods, except if trained on real data, are printed in \textbf{bold}.}
\label{tab2}
\begin{tabular}{l c c c c c c}
\hline
\multirow{2}{*}{\shortstack{\textbf{Classifier} \\ \textbf{training data}}} & \multicolumn{3}{c}{\textbf{PAT}} & \multicolumn{3}{c}{\textbf{HSI}}\\
\cline{2-7}
& BA & AUROC & F1-Score & BA & AUROC & F1-Score\\
\hline
\bfseries \textcolor{real}{\textbf{Real}} & 0.75 & 0.84 & 0.82 & 0.40 & 0.81 & 0.44\\
\hline
\bfseries \textcolor{simulated}{\textbf{Simulated}} & 0.52 & 0.64 & 0.64 & 0.24 & 0.75 & 0.18\\
\bfseries \textcolor{UNIT}{\textbf{UNIT}} & 0.50 & 0.44 & 0.65 & 0.20 & 0.72 & 0.20\\
\bfseries \textcolor{UNIT}{\textbf{$\text{UNIT}_\text{Y}$}} & 0.64 & \textbf{0.81} & 0.77 & 0.24 & 0.74 & \textbf{0.25}\\
\bfseries \textcolor{cINN}{\textbf{$\text{cINN}_\text{D}$}} & 0.66 & 0.73 & 0.72 & 0.25 & 0.72 & 0.20\\
\bfseries \textcolor{cINN}{\textbf{$\text{cINN}_\text{DY}$} (proposed)} & \textbf{0.71} & 0.80 & \textbf{0.78} & \textbf{0.29} & \textbf{0.76} & 0.24\\
\hline
\end{tabular}
\end{table}

\begin{table}[h!]
\centering
\caption{\textbf{Hyperparameters of cINN models and discriminators (Dis) for PAT and HSI.} The hyperparameters have been optimized to yield best spectral consistency and classification performance. The networks have been implemented in PyTorch. Here, ``lr'' represents the learning rate and ``WD'' the weight decay. The code and pre-trained models are available at: GITHUB\_LINK.}\label{tab1}
\begin{tabular}{l c c}
\hline
\bfseries Hyperparameter & \bfseries PAT cINN & \bfseries HSI cINN\\
\hline
\bfseries Epochs &  300 & 300\\
\bfseries Batch size &  2 & 10,000\\
\bfseries Optimizer & Adam (lr=0.001) & Adam (lr=0.0001) \\

\bfseries Optimizer parameters & $\beta_\text{1}$=0.4,$\beta_\text{2}$=0.999,WD=0.001 & $\beta_\text{1}$, $\beta_\text{2}$, WD=0.9, 0.95, 0.0001\\
\bfseries Scales & 5 & 1\\
\bfseries Blocks per scale & 4, 2, 1, 1, 2 & 40\\
\bfseries Conditions per scale & DY, D, D, D, D & 30$\times$DY + 10$\times$None\\
\bfseries Exponential clamping & 1 & 1\\
\shortstack{\bfseries Subnetwork\\ \,} & \shortstack{3$\times$(2D Conv + ReLU) \\ 256 hidden features} & \shortstack{2$\times$(Linear + ReLU) \\ 512 hidden features}\\
\hline
\bfseries Hyperparameter & \bfseries PAT Dis & \bfseries HSI Dis \\
\hline
\bfseries Scales & 3 & 1\\
\bfseries Layers per scale & 4$\times$(2D Conv + LeakyReLU) & 3$\times$(Linear + LeakyReLU)\\
\bfseries Hidden features & 64 & 256\\
\bfseries Dropout prob. & 0.2 & 0.2\\
\end{tabular}
\end{table}

\end{document}